\documentclass[useAMS,usenatbib,fleqn,a4paper]{mn2e}
\pdfoutput=1
\usepackage{times}
\usepackage{amsmath}
\usepackage{amssymb}
\usepackage{amsbsy}
\usepackage{graphicx}
\usepackage{subfigure}
\usepackage{paralist}
\usepackage{ mathrsfs }

\def\kpc{\,{\rm kpc}}

\def\pc{\,{\rm pc}}

\def\masyr{\,{\rm mas\,yr^{-1}}}
\def\kms{\,{\rm km\,s^{-1}}}
\def\rad{\,\rm rad}
\def\Gyr{\,{\rm Gyr}}
\def\d{{\rm d}}

\def\percent{\text{ per cent}}
\newcommand{\mat}[1]{\mathbfss{#1}}
\title[Tidal stream model in angle-frequency space]{Probabilistic model for constraining the Galactic potential using tidal streams}
\author[J. L. Sanders]{Jason L. Sanders\thanks{E-mail: jason.sanders@physics.ox.ac.uk}\\
Rudolf Peierls Centre for Theoretical Physics, Keble Road, Oxford OX1 3NP, UK}

\pagerange{\pageref{firstpage}--\pageref{lastpage}} \pubyear{2014}
\label{firstpage}
\begin{document}
\maketitle
\begin{abstract}

We present a generative probabilistic model for a tidal stream and demonstrate how this model is used to constrain the Galactic potential. The model takes advantage of the simple structure of a stream in angle and frequency space for the correct potential. We investigate how the method performs on full 6D mock stream data, and mock data with outliers included. As currently formulated the technique is computationally costly when applied to data with large observational errors, but we describe several modifications that promise to make the technique computationally tractable.
\end{abstract}


\section{Introduction}
A key goal in the study of the Milky Way is mapping the dark matter distribution of the Galaxy. Locally this is achieved through dynamical measurements of the disc components \citep{KuijkenGilmore1989,Garbari2012,BovyTremaine2012,Zhang2013}, whilst tidal streams present a tantalising prospect for constraining the dark matter distribution on a more global scale. These long, filamentary structures are the remnants of satellites tidally disrupted by the Milky Way. The stars in the stream are now essentially orbiting freely in the Galactic potential, but importantly they form a dynamically coherent structure, such that their properties are closely linked together. By inspecting the phase-space structure of tidal streams it should be possible to infer properties of the Galactic potential, and, in particular, the dark matter distribution.

Many methods have been proposed to achieve this. The simplest, and therefore most fruitful to-date, are orbit-fitting techniques \citep{Binney2008,EyreBinney2009B,EyreBinney2009A,Koposov2010,Deg2014}. Several authors have acknowledged the problems with simply fitting a stream with a single orbit \citep{EyreBinney2011,SandersBinney2013a,Lux2013}. Therefore, much work has been done on constructing methods for recovering the potential without assuming the stream delineates an orbit. One group of methods seeks the potential which minimises the spread in the integrals of motion \citep{Penarrubia2012}. Others allow for the expected range of orbits present in a stream given the progenitor properties \citep{Johnston1999,Varghese2011,PriceWhelan2013}.

The dynamics of tidal streams are very simply expressed in angle-action coordinates \citep{Tremaine1999,HelmiWhite1999,EyreBinney2011} denoted as $(\boldsymbol{\theta},\boldsymbol{J})$ throughout the paper. In \cite{SandersBinney2013b} we demonstrated the power of using the frequencies and angles of the stream to construct a measure of the goodness of fit of the potential. The frequencies, $\boldsymbol{\Omega}$, are the derivatives of the Hamiltonian, $H$, with respect to the actions i.e. $\boldsymbol{\Omega}=\partial H/\partial \boldsymbol{J}$. The best potential was deemed the one in which the angle and frequency distributions are aligned. This method suffered several drawbacks: we were doing the inference in model space, not the data space; it was difficult to assess the errors in the obtained potential parameters; it was awkward to handle errors in the data, and the method did not behave well for large errors.

Data for tidal streams have large errors and are expected to be contaminated with non-stream members. Streams lie at large distances from the Sun, far out in the halo. As such, the distance uncertainties can be significant, and small proper motion errors translate into large transverse velocity errors. For many streams we don't have full 6D data. Additionally, a stream is identified as a filament in the observable space -- often $l$ and $b$. Stream members are then extracted by cutting appropriately in these coordinates, and cleaning up by using the additional observables. This process obviously introduces many outliers to the stream data, which are perhaps just members of a smoother background halo, whilst also potentially throwing out stars which \emph{are} members of the stream.

This state of affairs leads us to analyse the data by constructing probabilistic models that correctly handle large errors, missing data, and contaminants. Such an approach is much more robust than previous efforts, and lends itself perfectly to being combined with other independent measurements of the Galactic potential. Here we present a probabilistic model for tidal streams that may be used to infer the properties of the Galactic potential. The model is expressed in the space of observables, but relies heavily on the expected structure of streams in angle-action space. In Section~\ref{Formalism} we motivate our choice of model by considering an idealised case of a Gaussian structure in angle-action space evolving in time. In Section~\ref{Model} we use these insights to write down a practical model for the stream, and discuss how it may be used to infer properties of the Galactic potential. In Section~\ref{Tests} we infer the parameters of a simple two-parameter potential from mock stream observations using our model. In Section~\ref{Discussion} we discuss proposed improvements to the approach taken in this paper.

\section{Formalism \& Motivation}\label{Formalism}
Given a set of observations, $D$, of $N$ stars believed to be members of a stream, what can we infer about the Galactic potential? For star $i$ we have a 6D set of Galactic coordinates $\boldsymbol{L}_i=(l,b,s,v_{||},\boldsymbol{\mu})$ with associated errors described by the covariance matrix $\mat{S}_i$. Note that we can fit any missing data into this formalism by taking the associated error with the data point to be infinite.

Given the data we want to know the posterior distribution of the potential, $\Phi$, given by $p(\Phi|D)$. From Bayes theorem we have
\begin{equation}
p(\Phi|D)=\frac{p(D|\Phi)p(\Phi)}{p(D)},
\end{equation}
where $p(\Phi)$ is the prior on the potential, and the evidence $p(D)$ is not important for the present exercise. We wish to evaluate the likelihood $p(D|\Phi)$.

The probability of the data given the potential is related to the properties of the stream progenitor, $\mathscr{C}$. $\mathscr{C}$ contains information about the current phase-space coordinates of the progenitor (i.e. the progenitor's actions, $\boldsymbol{J}_0$, frequencies, $\boldsymbol{\Omega}_0$, and angles, $\boldsymbol{\theta}_0$), as well as the size (and internal properties) of the progenitor. Therefore, we write
\begin{equation}
\begin{split}
&p(D|\Phi)=\int\mathrm{d}\mathscr{C}\,p(\mathscr{C})p(D|\Phi,\mathscr{C}),\\
&p(D|\Phi,\mathscr{C})=\prod_i^N p(\boldsymbol{L}_i|\Phi,\mathscr{C},\mat{S}_i),\\
&p(\boldsymbol{L}_i|\Phi,\mathscr{C},\mat{S}_i)=\int\mathrm{d}\boldsymbol{L}_i'\,p(\boldsymbol{L}_i|\boldsymbol{L}_i',\mat{S}_i)
\,\mathrm{det}\Big(\frac{\partial (\boldsymbol{x}_i,\boldsymbol{v}_i)}{\partial \boldsymbol{L}_i'}\Big)\times\\&\>\>\>\>\>\>\>\>\>\>\>\>\>\>\>\>\>\>\>\>\>\>\>\>\>\>\>\>\>\>\>\>\>\>\>\>\>\>\>\>\>\>\>\>p(\boldsymbol{x}_i,\boldsymbol{v}_i|\Phi,\mathscr{C}),
\end{split}
\label{Like}
\end{equation}
where
\begin{equation}
\begin{split}
p(\boldsymbol{L}_i|\boldsymbol{L}_i',\mat{S}_i)=&\frac{1}{\sqrt{(2\pi)^6\mathrm{det}(\mat{S}_i)}}\times\\&\exp\Big(-\frac{1}{2}(\boldsymbol{L}_{i}-\boldsymbol{L}_{i}')^{T}\mat{S}^{-1}(\boldsymbol{L}_{i}-\boldsymbol{L}_{i}')\Big),
\end{split}
\end{equation}
the Jacobian factor is given by
\begin{equation}
\mathrm{det}\Big(\frac{\partial (\boldsymbol{x}_i,\boldsymbol{v}_i)}{\partial \boldsymbol{L}_i'}\Big)=s'^4\cos b',
\end{equation}
and the ($\boldsymbol{x},\boldsymbol{v}$) coordinates are related to the Galactic coordinates in the usual way.

We want to work with actions, angles and frequencies. Throughout the paper we use the St\"ackel-fitting algorithm presented in \cite{Sanders2012} to estimate these quantities. This algorithm fits a St\"ackel potential to the region of the potential an orbit probes, and estimates the actions, angles and frequencies in the true potential as those in the best-fitting St\"ackel potential. Therefore, we write 
\begin{equation}
p(\boldsymbol{x}_i,\boldsymbol{v}_i|\Phi,\mathscr{C}) = \mathrm{det}\Big(\frac{\partial (\boldsymbol{\Omega}_i,\boldsymbol{\theta}_i)}{\partial (\boldsymbol{x}_i,\boldsymbol{v}_i)}\Big)p(\boldsymbol{\theta}_i,\boldsymbol{\Omega}_i|\Phi,\mathscr{C}),
\label{xv2angfreq}
\end{equation}
where the angles and frequencies are related to ($\boldsymbol{x},\boldsymbol{v}$) via the potential $\Phi$ using the St\"ackel-fitting approximation and the Jacobian is given by
\begin{equation}
\begin{split}
\mathrm{det}\Big(\frac{\partial (\boldsymbol{\Omega}_i,\boldsymbol{\theta}_i)}{\partial (\boldsymbol{x}_i,\boldsymbol{v}_i)}\Big) &= \mathrm{det}\Big(\frac{\partial (\boldsymbol{\Omega}_i,\boldsymbol{\theta}_i)}{\partial (\boldsymbol{J}_i,\boldsymbol{\theta}_i)}\Big)\mathrm{det}\Big(\frac{\partial (\boldsymbol{J}_i,\boldsymbol{\theta}_i)}{\partial (\boldsymbol{x}_i,\boldsymbol{v}_i)}\Big) \\&= \mathrm{det}\Big(\frac{\partial \boldsymbol{\Omega}_i}{\partial \boldsymbol{J}_i}\Big) = \mathrm{det}(\mat{D}_i),
\end{split}
\end{equation}
where we have used the fact that ($\boldsymbol{J},\boldsymbol{\theta}$) are canonical coordinates, such that the phase-space volume is conserved under the transformation, and introduced the Hessian matrix $\mat{D}$ defined as
\begin{equation}
\mat{D}\equiv\frac{\partial^2 H}{\partial \boldsymbol{J}^2}.
\label{Hessian}
\end{equation}
This matrix can be calculated analytically in a St\"ackel potential, so we extend the St\"ackel-fitting algorithm of \cite{Sanders2012} to estimate $\mat{D}$. We give details of this in the appendix. We proceed by splitting $p(\boldsymbol{\theta}_i,\boldsymbol{\Omega}_i|\Phi,\mathscr{C})$ into two components
\begin{equation}
p(\boldsymbol{\theta}_i, \boldsymbol{\Omega}_i | \Phi,\mathscr{C}) = p(\boldsymbol{\theta}_i| \boldsymbol{\Omega}_i, \Phi,\mathscr{C})p(\boldsymbol{\Omega}_i | \Phi,\mathscr{C}).
\end{equation}
To proceed further we must consider what we know about stream formation in angle-action space \citep{Tremaine1999}. Assuming the spread in actions in the stream is small \citep{SandersBinney2013a}, for each star in the stream we have
\begin{equation}
\begin{split}
&\Delta\boldsymbol{\Omega}_i=\boldsymbol{\Omega}_i-\boldsymbol{\Omega}_0\approx \mat{D}_0 \cdot (\boldsymbol{J}_i-\boldsymbol{J}_0) = \mat{D}_0 \cdot \Delta\boldsymbol{J}_i\\
&\Delta\boldsymbol{\theta}_i=\boldsymbol{\theta}_i-\boldsymbol{\theta}_0=t_i\Delta\boldsymbol{\Omega}_i+\Delta\boldsymbol{\theta}_i(0),
\end{split}
\label{StreamEquation}
\end{equation}
where $t_i$ is the time since the particle was stripped from the progenitor and $\Delta\boldsymbol{\theta}_i(0)$ is the separation between the $i$\textsuperscript{th} particle and the progenitor when the particle is released. $\mat{D}_0$ is the Hessian from equation~\eqref{Hessian} evaluated at the progenitor actions, $\boldsymbol{J}_0$.

To motivate our choice of model we begin by assuming that $\boldsymbol{J}_i$ follows an isotropic normal distribution such that
\begin{equation}
\begin{split}
p(\boldsymbol{J}_i | \Phi,\mathscr{C}) &\approx p(\boldsymbol{J}_i | \mathscr{C})\\
&=\frac{\sqrt{\mathrm{det}(\mat{A})}}{(2\pi)^\frac{3}{2}} \exp\Big(-\frac{1}{2}\Delta\boldsymbol{J}_i^T\cdot \mat{A} \cdot\Delta\boldsymbol{J}_i\Big) \\
&= \Big(\frac{a}{2\pi}\Big)^\frac{3}{2}\exp\Big(-\frac{a}{2} |\Delta\boldsymbol{J}_i|^2\Big),
\end{split}
\end{equation}
where $a$ gives the spread of the action distribution. This is related to the progenitor mass, $M$, by $a\propto M^{2/3}$ \citep{SandersBinney2013a}. Such a simple model for the action distribution is unrealistic \citep{EyreBinney2011} but our understanding of this simplistic model will aid in the construction of a more realistic model. Similarly we assume that $\Delta\boldsymbol{\theta}_i(0)$ is distributed as an isotropic Gaussian such that
\begin{equation}
p(\Delta\boldsymbol{\theta}_i(0))=\Big(\frac{b}{2\pi}\Big)^\frac{3}{2}\exp\Big(-\frac{b}{2}|\Delta\boldsymbol{\theta}_i(0)|^2\Big).
\label{ptheta0}
\end{equation}
This Gaussian model for a stream in actions and initial angles was studied by \cite{HelmiWhite1999}.
From equation~\eqref{StreamEquation} the frequency is linearly related to the actions via the Hessian, $\mat{D}_0$, so we can write down the distribution for the frequencies as
\begin{equation}
\begin{split}
p(\boldsymbol{\Omega}_i | \Phi,\mathscr{C})&=\mathrm{det}\Big(\frac{\partial \boldsymbol{J}_i}{\partial \boldsymbol{\Omega}_i}\Big)p(\boldsymbol{J}_i | \Phi,\mathscr{C})\\
&\approx\mathrm{det}(\mat{D}^{-1}_0)\Big(\frac{a}{2\pi}\Big)^\frac{3}{2} \exp\Big(-\frac{a}{2} \Delta\boldsymbol{\Omega}_i^T \mat{D}_0^{-1}\mat{D}_0^{-1}\Delta\boldsymbol{\Omega}_i\Big),
\end{split}
\label{FreqDist}
\end{equation}
where as the spread in actions is small we have approximated the Jacobian by its value at the progenitor actions.

This distribution is a multivariate normal distribution with principal axes along the principal eigenvectors of $\mat{D}_0$ and with width given by the corresponding eigenvalues. $\mat{D}_0$ is a symmetric matrix so has real eigenvalues and orthogonal eigenvectors. Note here that for long thin streams to form $\mat{D}_0$ has one eigenvalue much greater than the other two. In \cite{SandersBinney2013a} we demonstrated that in a realistic Galactic potential this condition was satisfied for a large volume of action space. Therefore we write
\begin{equation}
\begin{split}
\mat{D}_0^{-1} &= \sum_j^3 \frac{1}{\lambda_j}\hat{\boldsymbol{e}}_j\cdot\hat{\boldsymbol{e}}_j^T\\
&\approx \frac{1}{\lambda_2}\hat{\boldsymbol{e}}_2\cdot\hat{\boldsymbol{e}}_2^T+\frac{1}{\lambda_3}\hat{\boldsymbol{e}}_3\cdot\hat{\boldsymbol{e}}_3^T
\end{split}
\end{equation}
where $\lambda_j$ and $\hat{\boldsymbol{e}}_j$ are the eigenvalues and eigenvectors of $\mat{D}_0^{-1}$, and we have $\lambda_1\gg\lambda_2>\lambda_3$. Therefore we find that 
\begin{equation}
\begin{split}
p(\boldsymbol{\Omega}_i | \Phi,\mathscr{C})\propto\exp\Big(-\frac{a}{2\lambda_2^2} (\Delta\boldsymbol{\Omega}_i\cdot\hat{\boldsymbol{e}}_2)^2-\frac{a}{2\lambda_3^2} (\Delta\boldsymbol{\Omega}_i\cdot\hat{\boldsymbol{e}}_3)^2\Big).
\label{FinalFreq}
\end{split}
\end{equation}
The distribution of frequencies is 2D Gaussian perpendicular to a straight line in frequency space defined by $\hat{\boldsymbol{e}}_1$. In the simple model presented here the distribution along $\hat{\boldsymbol{e}}_1$ is also Gaussian. However, we will later adopt a superior distribution along $\hat{\boldsymbol{e}}_1$ which better reflects the stream distribution.

Next we address the angle distribution. The angles depend upon the additional variables, $t_i$ and $\Delta\boldsymbol{\theta}_i(0)$. Therefore, we write
\begin{equation}
\begin{split}
p(\boldsymbol{\theta}_i| \boldsymbol{\Omega}_i, \Phi,\mathscr{C}) = \int \mathrm{d}t_i&\,\mathrm{d}^3\Delta\boldsymbol{\theta}_i(0)\,p(\boldsymbol{\theta}_i| \boldsymbol{\Omega}_i, \Delta\boldsymbol{\theta}_i(0), t_i, \mathscr{C})\times\\&p(t_i)p(\Delta\boldsymbol{\theta}_i(0)).
\label{ptheta}
\end{split}
\end{equation} 
Given a time since stripping, a frequency separation, and an initial angle separation, the present angle separation is completely determined by equation~\eqref{StreamEquation} so
\begin{equation}
p(\boldsymbol{\theta}_i| \boldsymbol{\Omega}_i, \Delta\boldsymbol{\theta}_i(0), t_i, \mathscr{C}) = \delta^3(\Delta\boldsymbol{\theta}_i-t_i\Delta\boldsymbol{\Omega}_i-\Delta\boldsymbol{\theta}_i(0)).
\end{equation}
Substituting this and equation~\eqref{ptheta0} into equation~\eqref{ptheta} and performing the integral over $\Delta\boldsymbol{\theta}_i(0)$ using the $\delta$-function we have that
\begin{equation}
\begin{split}
p(\boldsymbol{\theta}_i| \boldsymbol{\Omega}_i, \Phi,\mathscr{C}) = &\int \mathrm{d}t_i\,p(t_i)\Big(\frac{b}{2\pi}\Big)^\frac{3}{2} \times\\&\exp\Big(-\frac{b}{2}|\Delta\boldsymbol{\theta}_i-t_i\Delta\boldsymbol{\Omega}_i|^2\Big).\end{split}
\label{pthetat}
\end{equation}
Now we rearrange the argument of the exponential as
\begin{equation}
\begin{split}
|\Delta&\boldsymbol{\theta}_i-t_i\Delta\boldsymbol{\Omega}_i|^2 \\
&=|\Delta\boldsymbol{\Omega}_i|^2\Big(t_i-\frac{\Delta\boldsymbol{\theta}_i\cdot\Delta\boldsymbol{\Omega}_i}{|\Delta\boldsymbol{\Omega}_i|^2}\Big)^2-\frac{(\Delta\boldsymbol{\theta}_i\cdot\Delta\boldsymbol{\Omega}_i)^2}{|\Delta\boldsymbol{\Omega}_i|^2}+|\Delta\boldsymbol{\theta}_i|^2,
\end{split}
\end{equation}
and note that $\Delta\boldsymbol{\Omega}_i \approx \lambda_1\hat{\boldsymbol{e}}_1(\hat{\boldsymbol{e}}_1\cdot \Delta\boldsymbol{J}_i)$ so
\begin{equation}
\frac{(\Delta\boldsymbol{\theta}_i\cdot\Delta\boldsymbol{\Omega}_i)^2}{|\Delta\boldsymbol{\Omega}_i|^2} \approx (\Delta\boldsymbol{\theta}_i\cdot\hat{\boldsymbol{e}}_1)^2,
\end{equation}
and
\begin{equation}
-\frac{(\Delta\boldsymbol{\theta}_i\cdot\Delta\boldsymbol{\Omega}_i)^2}{|\Delta\boldsymbol{\Omega}_i|^2}+|\Delta\boldsymbol{\theta}_i|^2 \approx (\Delta\boldsymbol{\theta}_i\cdot\hat{\boldsymbol{e}}_2)^2+(\Delta\boldsymbol{\theta}_i\cdot\hat{\boldsymbol{e}}_3)^2.
\end{equation}
Therefore, equation~\eqref{pthetat} becomes
\begin{equation}
\begin{split}
p(\boldsymbol{\theta}_i| \boldsymbol{\Omega}_i&, \Phi,\mathscr{C}) \approx \Big(\frac{b}{2\pi}\Big)^\frac{3}{2} \exp\Big(-\frac{b}{2}\sum_{k=2,3}(\Delta\boldsymbol{\theta}_i\cdot\hat{\boldsymbol{e}}_k)^2\Big)\\\times
& \int \mathrm{d}t_i\,p(t_i)\exp\Big(-\frac{b|\Delta\boldsymbol{\Omega}_i|^2
}{2}\Big(t_i-\frac{\Delta\boldsymbol{\theta}_i\cdot\hat{\boldsymbol{e}}_1}{\Delta\boldsymbol{\Omega}_i\cdot\hat{\boldsymbol{e}}_1}\Big)^2\Big).
\end{split}
\label{FinalAng}
\end{equation}
The first part is a 2D Gaussian perpendicular to the eigenvector $\hat{\boldsymbol{e}}_1$ (as with the frequencies) whilst the second part depends on when the particles were stripped from the progenitor and only affects the angle distribution along the vector $\hat{\boldsymbol{e}}_1$ i.e. ($\Delta\boldsymbol{\theta}_i\cdot\hat{\boldsymbol{e}}_1$). If we assume that $p(t_i)$ is uniform (see below for discussion) and $\Delta\boldsymbol{\theta}\gg\Delta\boldsymbol{\theta}(0)$ (i.e. the particle was stripped long enough ago that the time part of equation~\eqref{StreamEquation} dominates the initial angle separation from the progenitor -- this is an assumption we made in \cite{SandersBinney2013b}), we may perform the $t_i$ integral to find
\begin{equation}
\begin{split}
p(\boldsymbol{\theta}_i| \boldsymbol{\Omega}_i, \Phi,\mathscr{C}) \approx &\frac{b}{2\pi|\Delta\boldsymbol{\Omega}_i|t_{\rm max}}\exp\Big(-\frac{b}{2}\sum_{k=2,3}(\Delta\boldsymbol{\theta}_i\cdot\hat{\boldsymbol{e}}_k)^2\Big)\\
&\text{ if }0<\frac{\Delta{\boldsymbol{\theta}}_i\cdot\hat{\boldsymbol{e}}_1}{\Delta{\boldsymbol{\Omega}}_i\cdot\hat{\boldsymbol{e}}_1}<t_{\rm max},
\end{split}
\label{UniformTime}
\end{equation}
where the condition ensures that the stripping time for each stream member is positive, and less than some maximum stripping time, $t_{\rm max}$. This expression demonstrates explicitly that the distribution perpendicular to $\hat{\boldsymbol{e}}_1$ is independent of the distribution along $\hat{\boldsymbol{e}}_1$. In conclusion, in this model both the angle and frequency distributions are highly elongated along the vector $\hat{\boldsymbol{e}}_1$. This validates the procedure followed in \cite{SandersBinney2013b}.

The assumption of a uniform stripping-time distribution, $p(t_i)$, does not well model the highly-concentrated stripping events around pericentric passage observed in N-body simulations of clusters on eccentric orbits. However, as shown in equation~\eqref{FinalAng} the exact form adopted for $p(t_i)$ only affects the density of particles along the stream i.e. the structure of the angle distribution along $\hat{\boldsymbol{n}}$. For diagnosis of the Galactic potential and mass distribution we are interested in the shape of the stream, so the real diagnostic power comes from the clumping in frequency space and the alignment of the frequency and angle distributions. Our simple assumption of uniform stripping times should not affect the recovery of the potential parameters significantly. We will see in \S~\ref{Tests} that $p(t_i)$ for a stream generated from an N-body simulation is not uniform. However, the potential parameters are recovered successfully using this stream data when the assumption of a uniform stripping-time distribution is made.

\section{Model}\label{Model}
In the formalism of the previous section we made several assumptions that, while useful for illustrative purposes, we would like to relax. The assumption of isotropic $\Delta\boldsymbol{J}$ distribution is not valid as evidenced in \cite{EyreBinney2011}. For a general action distribution we would still expect a highly anisotropic frequency distribution but the principal eigenvector of this distribution will not be that of the Hessian matrix $\mat{D}$, but some other vector $\hat{\boldsymbol{n}}$, with vectors $\hat{\boldsymbol{d}}_1$ and $\hat{\boldsymbol{d}}_2$ perpendicular to this. The intricacies of the action distribution will be reflected in the frequency distribution along the vector $\hat{\boldsymbol{n}}$. Additionally the angle distribution will also be highly elongated along this direction $\hat{\boldsymbol{n}}$. Analogous to a combination of equation~\eqref{FinalFreq} and \eqref{FinalAng} we write 
\begin{equation}
\begin{split}
p(\boldsymbol{\theta}_i, \boldsymbol{\Omega}_i | \Phi,\mathscr{C}) &= \frac{K_\theta(\boldsymbol{\theta}_i|\boldsymbol{\Omega}_i)}{2\pi u^2}\exp\Big[-\sum_{j=1,2}\frac{1}{2u^2}(\boldsymbol{\theta}_i\cdot\hat{\boldsymbol{d}}_j-\gamma_j)^2\Big] \\		&\times\frac{K_\Omega(\boldsymbol{\Omega}_i)}{2\pi w^2}\exp\Big[-\sum_{j=1,2}\frac{1}{2w^2}(\boldsymbol{\Omega}_i\cdot\hat{\boldsymbol{d}}_j-\omega_j)^2\Big],
\end{split}
\label{pFreqAngFinal}
\end{equation}
where the functions $K_i$ define the stream distribution along the vector $\hat{\boldsymbol{n}}$ in the angle or frequency space. The quantities $u$ and $w$ are the widths perpendicular to this vector in angle and frequency space, respectively. Note we have assumed the stream is isotropically distributed perpendicular to the vector $\hat{\boldsymbol{n}}$. $\omega_j$ and $\gamma_j$ are related to the present frequency and angle coordinates of the progenitor.

We define the angles $\phi$ and $\psi$ such that
\begin{equation}
\hat{\boldsymbol{n}}=(\sin\phi\cos\psi,\cos\phi\cos\psi,\sin\psi),
\end{equation}
and we choose
\begin{equation}
\begin{split}
\hat{\boldsymbol{d}}_1&=(\cos\phi,-\sin\phi,0),\\
\hat{\boldsymbol{d}}_2&=(\sin\phi\sin\psi,\cos\phi\sin\psi,-\cos\psi).
\end{split}
\end{equation}
Note this choice of vectors perpendicular to $\hat{\boldsymbol{n}}$ is arbitrary. We have set the distribution perpendicular to $\hat{\boldsymbol{n}}$ to be isotropic so our choice of vectors is unimportant. We now specify the functions $K_i$ defining the stream distribution along the vector $\hat{\boldsymbol{n}}$. In frequency space the distribution along $\hat{\boldsymbol{n}}$ consists of two separated peaks corresponding to the leading and trailing tails of the stream (see next section). For simplicity we assume that each of these peaks is Gaussian. The angle distribution depends upon both the frequency distribution and the distribution of stripping times. As in equation~\eqref{UniformTime}, we make the simple first-order assumption of a uniform stripping-time distribution such that the distribution in angle space along the stream given a frequency separation is also uniform between $0$ and some maximum stripping time, $t_{\rm max}$. Therefore, we write
\begin{equation}
\begin{split}
&K_\Omega(\boldsymbol{\Omega}_i) = \frac{1}{\sqrt{2\pi w_0^2}}\sum_{k=\pm 1}\exp\Big[-\frac{1}{2w_0^2}(\boldsymbol{\Omega}_i\cdot\hat{\boldsymbol{n}}-\omega_0+k\Omega_s)^2\Big],\\
&K_\theta(\boldsymbol{\theta}_i|\boldsymbol{\Omega}_i) = 
\begin{cases}
\frac{1}{|\boldsymbol{\Omega}_i\cdot\hat{\boldsymbol{n}}-\omega_0|t_{\rm max}},& \text{if }0<\frac{(\boldsymbol{\theta}_i\cdot\hat{\boldsymbol{n}}-\gamma_0)}{(\boldsymbol{\Omega}_i\cdot\hat{\boldsymbol{n}}-\omega_0)}<t_{\rm max}, \\
    0,              & \text{otherwise}.
\end{cases}
\end{split}
\end{equation}
$2\Omega_s$ gives the separation between the Gaussian peaks along $\hat{\boldsymbol{n}}$ in frequency space. When equation~\eqref{pFreqAngFinal} is combined with equation~\eqref{Like} and \eqref{xv2angfreq} we have completely specified our model. Given a set of 6D stream data with associated errors we can assess the likelihood of a given potential by evaluating the integral of equation~\eqref{Like}. It is defined by 13 progenitor parameters given by $\mathscr{C}=\{\phi,\psi,\gamma_j,\omega_j,u,w,w_0,t_{\rm max},\Omega_s\}$ and $\mathscr{N}$ potential parameters.

\subsection{MCMC}
We sample from the posterior using Markov chain Monte Carlo. We use an affine-invariant sampler implemented in the \textit{emcee} package from \cite{ForemanMackey2013}. For each of the following tests we use a group of $144$ walkers, and vary the nuisance parameters $\mathscr{C}$ as well as the potential parameters. For all scale parameters (i.e. $u,w,w_0,t_{\rm max}$) we use a logarithmic flat prior, whilst for the other parameters we use uniform flat priors.

To perform the integral over the errors in the calculation of the likelihood we use the Vegas Monte Carlo integration algorithm \citep{Lepage} implemented in the Gnu Science Library \citep{GSL}. Our stream model is typically very narrow whilst the error distribution for each observable coordinate can be very broad. Therefore there is a very small region of the 4D integrand which has any support. Using an adaptive integration scheme such as Vegas means we can rapidly focus on this small region.

\section{Tests}\label{Tests}
\begin{figure}
$$\includegraphics[bb = 7 7 209 412]{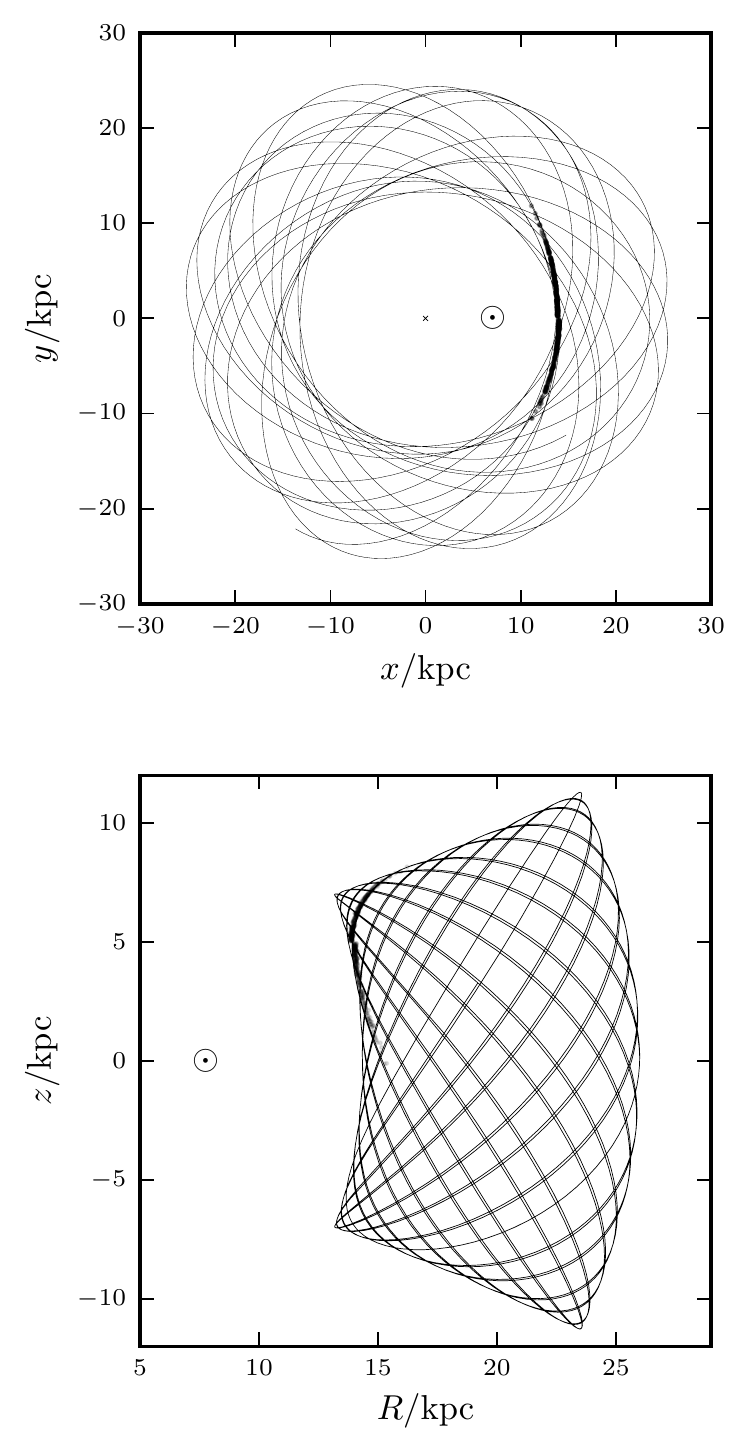}$$
\caption{Stream simulation in real space -- the stream particles are shown in black with the 30 selected particles shown in red, and the progenitor orbit is given by the black line. The Sun is marked by $\odot$, and the black cross on the left panel is the Galactic Centre.}
\label{Orbit}
\end{figure}

We test the above procedure using particles taken from a stream simulation. The potential we work with is the two-parameter ($\mathscr{N}=2$) logarithmic potential given by
\begin{equation}
\Phi(R,z) = \frac{V_c^2}{2}\log\Big(R^2+\frac{z^2}{q^2}\Big).
\end{equation}
The two parameters of this potential are $V_c$ and $q$. We set $V_c=220\kms$ and $q=0.9$ for the simulation. We place a cluster at the apocentre of the orbit shown in Fig.~\ref{Orbit}. This orbit was chosen due to its similarity to the orbit of the GD-1 stream \citep{Koposov2010}. The orbit has initial conditions $(R,z)=(26.0,0.0)\kpc$ and $(U,V,W)=(0.0,141.8,83.1)\kms$ where positive $U$ is towards the Galactic centre and positive $V$ is in the direction of the Galactic rotation at the Sun.  We seed the simulation with $10000$ particles drawn from a $2\times 10^4M_\odot$ King profile with the ratio of central potential to squared-velocity parameter, $W_0 = \Psi_0/\sigma^2 = 2$, a tidal limiting radius, $r_t\approx70\pc$, and a King core radius of $\approx20\pc$. We evolve the simulation for $t\approx4.2\Gyr$ (until just after $11$\textsuperscript{th} pericentre passage of the progenitor) using the code {\textsc{gyrfalcON}} \citep{Dehnen2000,Dehnen2002}. 

\begin{figure}
$$\includegraphics[bb = 7 7 205 223]{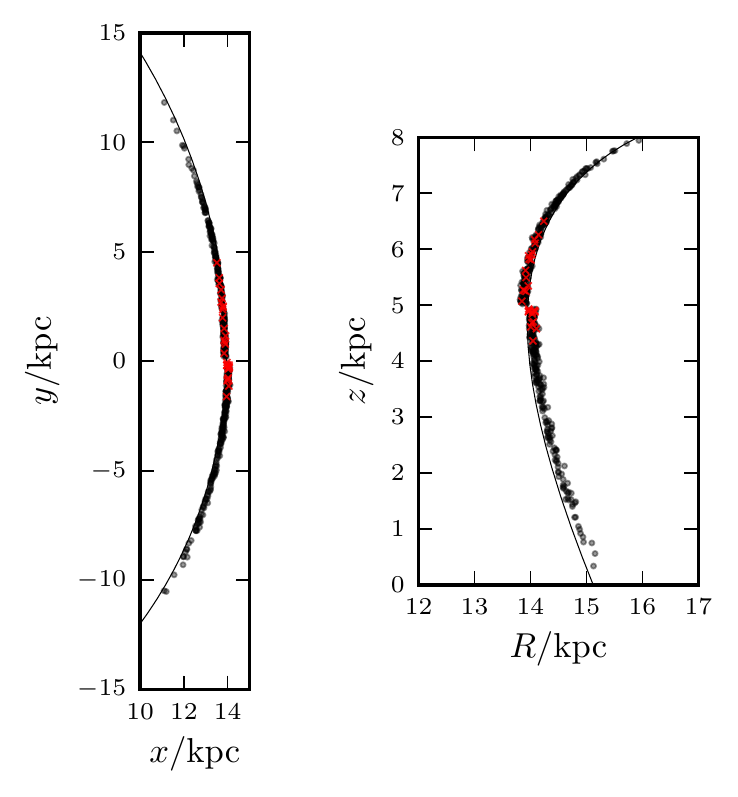}$$
\caption{Stream simulation in real space -- the stream particles are shown in black with the 30 selected particles shown in red, and the progenitor orbit is given by the black line. The Sun is marked by $\odot$, and the black cross on the left panel is the Galactic Centre.}
\label{ZoomInOrbit}
\end{figure}

\begin{figure}
$$\includegraphics[bb = 7 7 231 226]{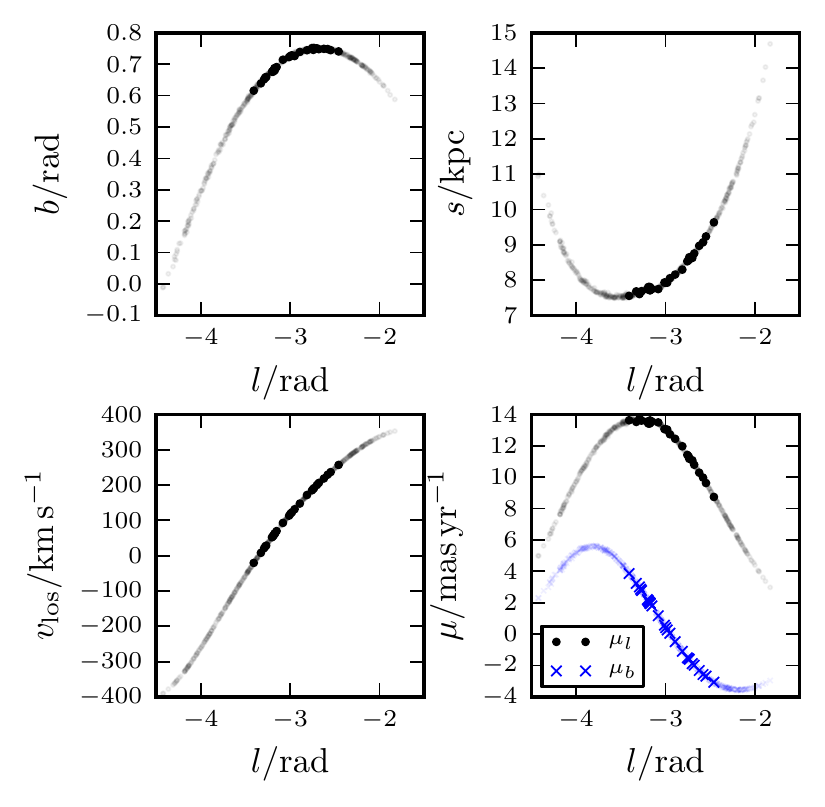}$$
\caption{Stream particles in galactic coordinates with the 30 selected particles indicated by the heavy markers.}
\label{galcoords}
\end{figure}

We take the resulting distribution of particles, remove the progenitor remnant with a spatial cut, and rotate the coordinate frame such that the Sun is placed at the same azimuthal angle as the progenitor. From the resulting particles we randomly select $30$ particles that lie in the range $-200^\circ<l<-140^\circ$. The chosen sample of particles are shown in Fig.~\ref{ZoomInOrbit} and Fig.~\ref{galcoords}. For the tests shown below we include the observational errors by scattering the observed coordinates by the appropriate gaussian errors.

\begin{figure}
$$\includegraphics[bb = 7 7 164 230]{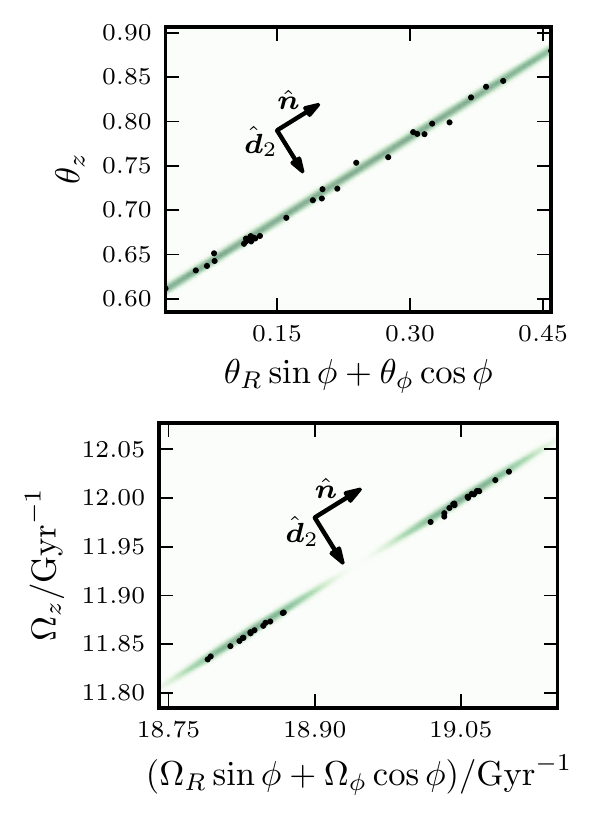}$$
\caption{Mock data and model in angle-frequency space  -- the black points give the angles and frequencies of the 30 selected particles calculated using the correct potential. The green shows the logarithm of the value of the model given in equation~\eqref{pFreqAngFinal} on a plane through angle-frequency space which contains the vectors $\hat{\boldsymbol{n}}$ and $\hat{\boldsymbol{d}}_2$. We see that in both angle and frequency space the stream has a linear structure aligned with the vector $\hat{\boldsymbol{n}}$.}
\label{model}
\end{figure}

\begin{figure}
$$\includegraphics[bb = 7 7 221 105]{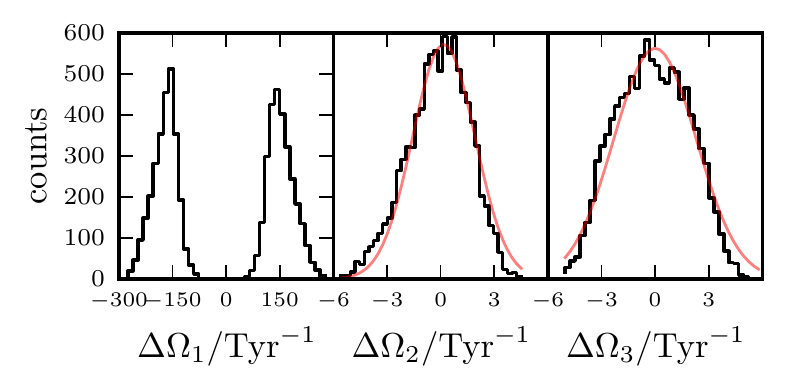}$$
\caption{Histograms of the frequencies for \emph{all} stream particles projected along the vectors $\hat{\boldsymbol{n}}$, $\hat{\boldsymbol{d}}_1$ and $\hat{\boldsymbol{d}}_2$ such that $\Delta\Omega_1 = (\Delta\boldsymbol{\Omega}\cdot\hat{\boldsymbol{n}})$, $\Delta\Omega_2 = (\Delta\boldsymbol{\Omega}\cdot\hat{\boldsymbol{d}}_1)$, and $\Delta\Omega_3 = (\Delta\boldsymbol{\Omega}\cdot\hat{\boldsymbol{d}}_2)$. The middle and right panels show Gaussian fits to the histograms in red.}
\label{distsperp}
\end{figure}

In Fig.~\ref{model} we plot the angles and frequencies in the correct potential, along with a cross-section through the model specified in equation~\eqref{pFreqAngFinal}. The parameters of the model were chosen as those that maximised the likelihood in the correct potential. We see that in the correct potential the angle and frequency structure of the stream takes on a simple linear distribution. In Fig.~\ref{distsperp} we plot the projections of the frequency histograms along the vectors $\hat{\boldsymbol{n}}$, $\hat{\boldsymbol{d}}_1$ and $\hat{\boldsymbol{d}}_2$ for all particles in the stream from the simulation. We see that the distribution perpendicular to $\hat{\boldsymbol{n}}$ is approximately Gaussian. The distribution along $\hat{\boldsymbol{n}}$ consists of two peaks consisting of the leading and trailing tail. Each peak is skewed such that there is a longer tail towards larger values of $|\Delta\boldsymbol{\Omega}\cdot\hat{\boldsymbol{n}}|$. The structure of these peaks was discussed by \cite{Johnston1998}. 

\begin{figure}
$$\includegraphics[bb = 7 7 222 152]{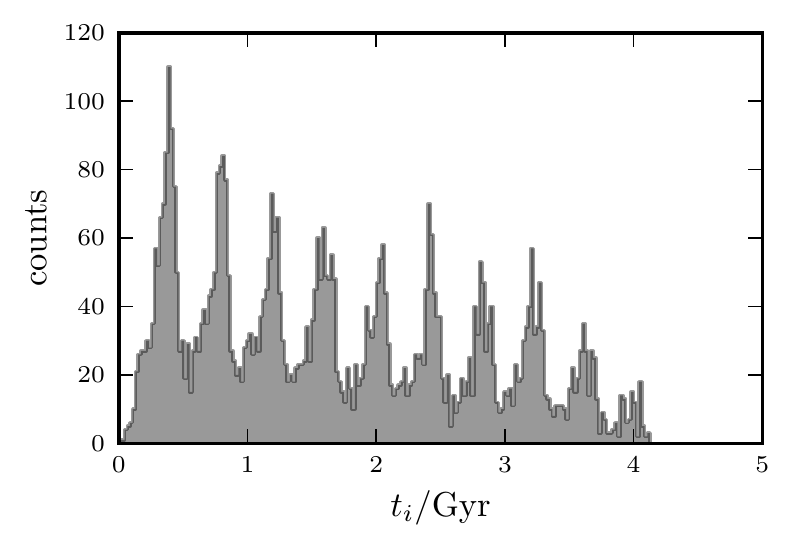}$$
\caption{Time since release for the particles in the stream estimated as $t_i=|\Delta\btheta_i|/|\Delta\boldsymbol{\Omega}_i|$ in the true potential. The distribution is peaked around pericentric passage, and in general more particles are stripped at later times.}
\label{releasetimes}
\end{figure}

Our model assumes a uniform stripping-time distribution. In Figure~\ref{releasetimes} we plot the time since release for the particles in the stream estimated as $t_i=|\Delta\btheta_i|/|\Delta\boldsymbol{\Omega}_i|$ found in the true potential. We see that the distribution is peaked around pericentric passage with slightly more particles being stripped at later times as the cluster mass decreases. If the stream has undergone several stripping events then, if we average on a timescale comparable to the radial period, the distribution of stripping times is approximately uniform. As mentioned previously, we expect the assumption of a uniform stripping-time distribution to be appropriate for measuring the Galactic potential, but more detailed modelling is required if we wish to reproduce the density distribution along the stream.

By only sampling a portion of the stream track on the sky we have limited the range of available angles of the stream particles, but we expect that the range of frequencies sampled is fair. We will miss some high frequency separations which are stripped earliest. In this situation the $t_{\rm max}$ parameter tells us about the time since the first of the observed particles were stripped ($\sim 2\Gyr$). 

\subsection{No contaminants}
\begin{figure}
$$\includegraphics[bb = 7 7 260 196]{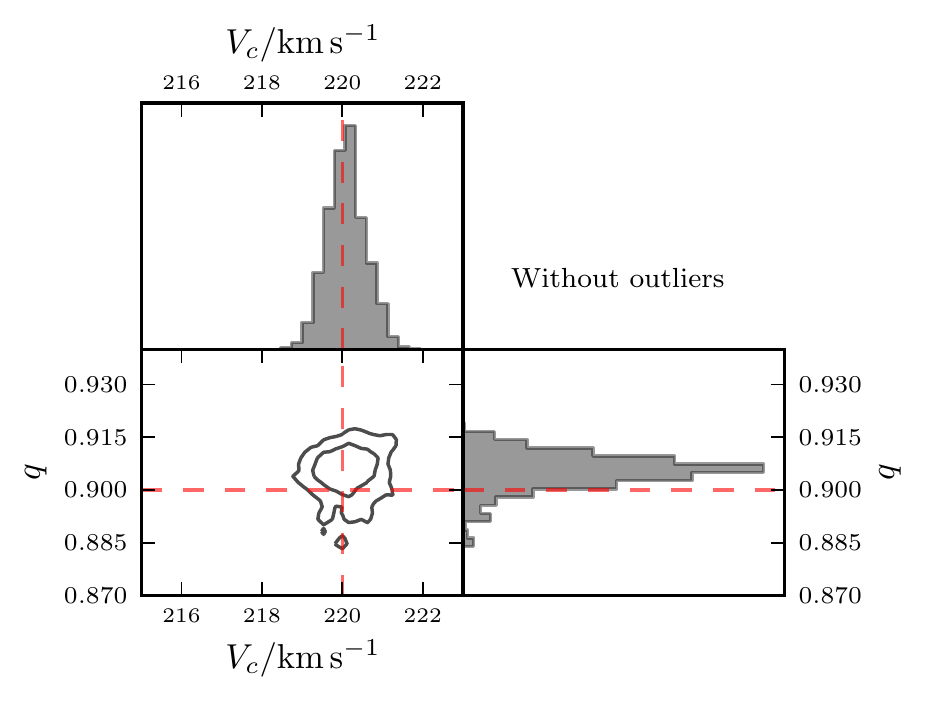}$$
\caption{Posterior distributions of the potential parameters for error-free data. The contours contain $68\percent$ and $95\percent$ of the samples from an MCMC chain. The red dashed lines show the parameters used to produce the simulation.}
\label{errfree}
\end{figure}

We test the above method by considering a dataset where each of our stream particles has full 6D data, $(l,b,s,v_{||},\boldsymbol{\mu})$. We first demonstrate that the method works for error-free data. In Figure~\ref{errfree} we show the posterior distributions of the potential parameters. The correct potential is recovered with $\sim 0.5\kms$ and $0.005$ errors in $V_c$ and $q$ respectively. 

\begin{figure}
$$\includegraphics[bb = 7 7 260 197]{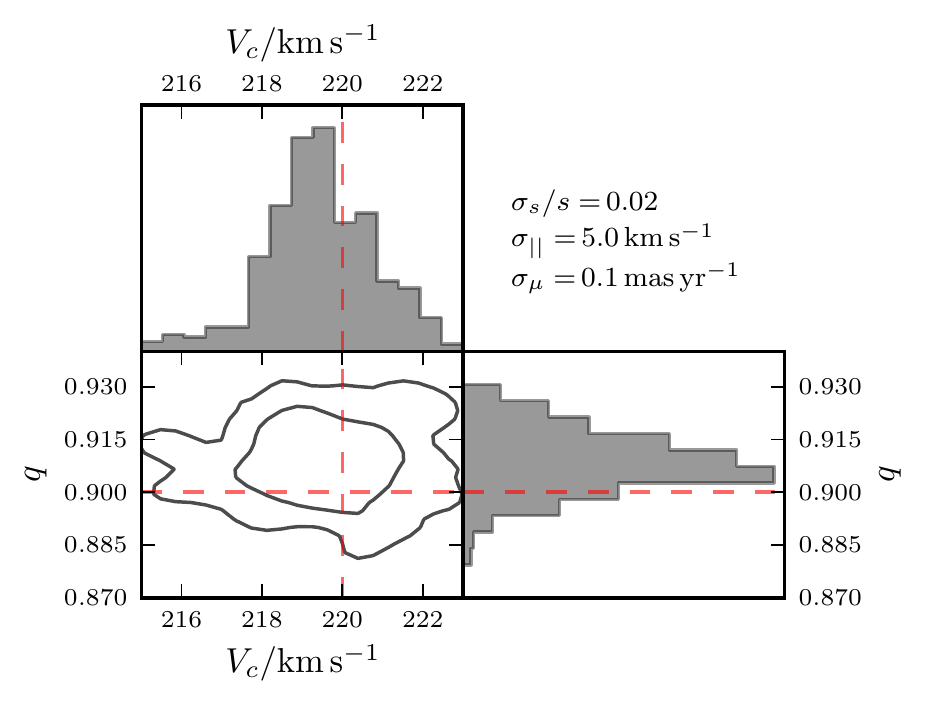}$$
\caption{Posterior distributions of the potential parameters for data with the errors shown in the top right corner. The black contours enclose $68\percent$ and $95\percent$ of the samples. The red dashed lines show the parameters used to produce the simulation.}
\label{vegas}
\end{figure}

Next, we consider an \emph{optimistic} dataset where we assume each of the stars is an RR Lyrae observed by Spitzer \citep{PriceWhelan2013}. We assume that the covariance matrix is diagonal and identical for all data such that $S_{jk} = \delta_{jk}\sigma_j^2$. We adopt $2\percent$ distance errors ($\sigma_s/s$), $5\kms$ line-of-sight velocity errors ($\sigma_{||}$), and $0.1\masyr$ proper motion errors ($\sigma_\mu$). Such a dataset is unrealistic for such a low mass stream. However, it is suitable for demonstrating the method. We fix the parameters $w_0=0.08\Gyr^{-1}$, $u=0.02\rad$ and $w=0.006\Gyr^{-1}$, and let the others vary as before. This prior essentially sets the mass of the cluster. The posterior distribution of the recovered potential parameters is shown in Figure~\ref{vegas}. We recover the correct potential parameters as $V_c=(219.4\pm1.4)\kms$ and $q=(0.909\pm0.009)$ approximating the posterior as an uncorrelated Gaussian.

\begin{figure}
$$\includegraphics[bb = 7 7 239 226]{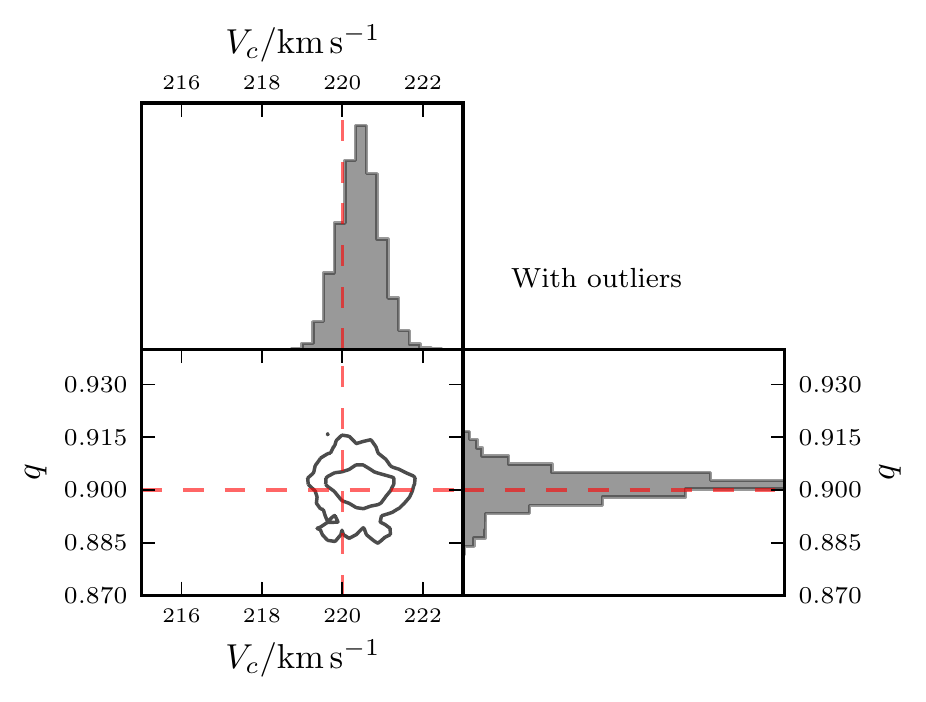}$$
\caption{Stream and outlier particles in galactic coordinates.}
\label{galcoords_outlier}
\end{figure}

\begin{figure}
$$\includegraphics[bb = 7 7 260 196]{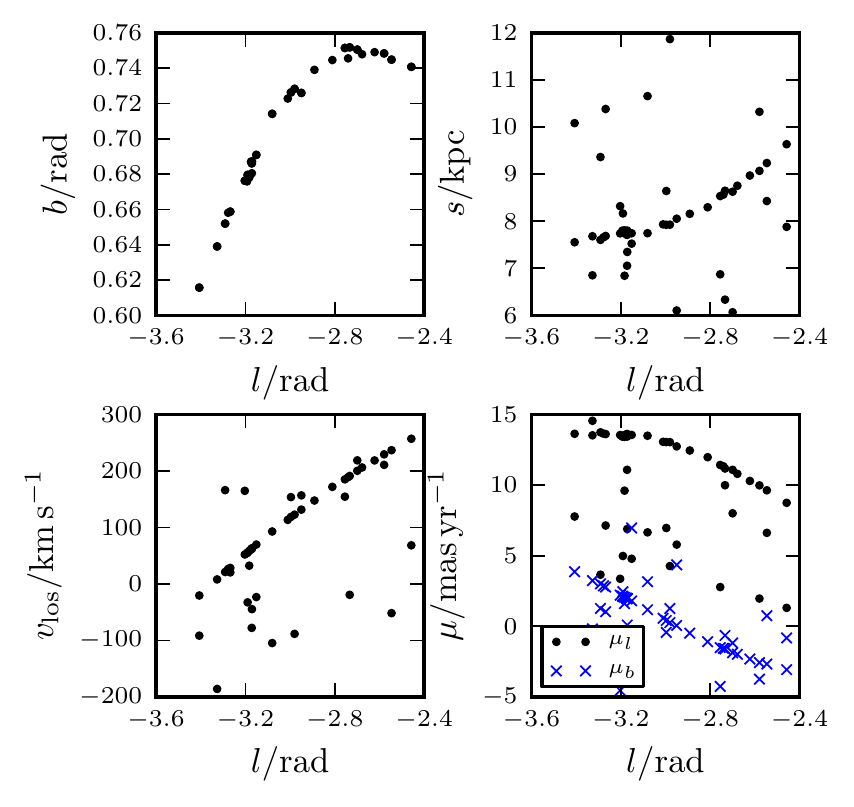}$$
\caption{Posterior distributions of the potential parameters for error-free data with outliers included. The contours contain $68\percent$ and $95\percent$ of the samples from an MCMC chain. The red dashed lines show the parameters used to produce the simulation.}
\label{errfreeoutlier}
\end{figure}

\begin{figure}
$$\includegraphics[bb = 7 7 232 174]{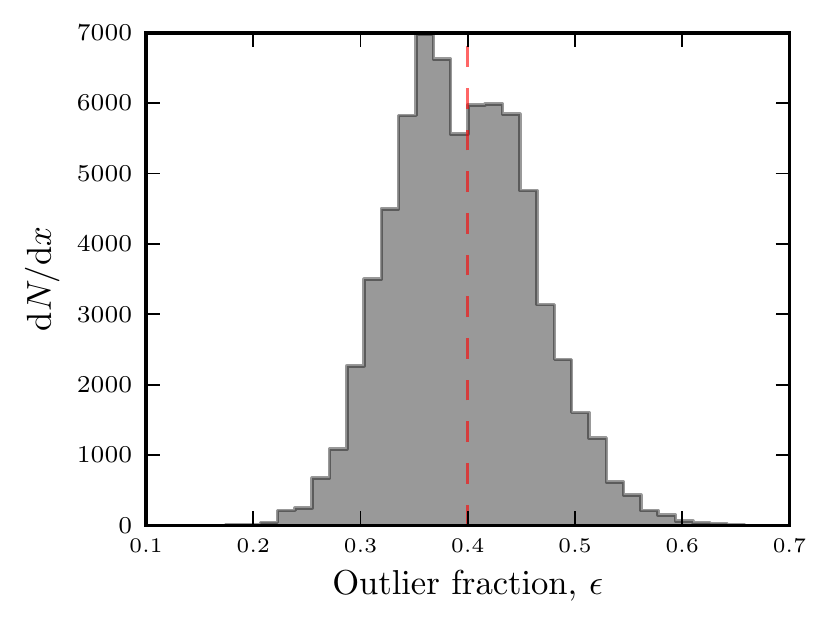}$$
\caption{Posterior distribution for the outlier fraction for the error-free dataset. The red dashed lines show the outlier fraction used to produce the mock data.}
\label{outlier_frac}
\end{figure}

\subsection{Inclusion of outliers}
Stream data are inevitably contaminated with stars that are not associated with the stream. Many authors attempt to remove these outliers by performing cuts in the observable space. However, it is much better to model the outliers. To include $m$ outliers in our test we randomly select $m$ stars from our simulation sample. We use the $l$ and $b$ values of these $m$ stars for our outlier stars. For each star we sample a distance, $s$, from a uniform distribution between $6\kpc$ and $12\kpc$. We convert the tuple $(l,b,s)$ to a Cartesian position, $\boldsymbol{x}$. At this position $\boldsymbol{x}$ we draw a set of velocities $\boldsymbol{v}$ from an isothermal distribution function given by
\begin{equation}
	p_{\rm iso}(\boldsymbol{x}_i,\boldsymbol{v}_i|\Phi) = p_h(E|\Phi)\propto \exp (-E/\sigma_h^2),
\end{equation}
with dispersion $\sigma_h=100\mathrm{km\,s}^{-1}$.

In our model we assume that the likelihood of a star at angle-frequency coordinates $(\boldsymbol{\Omega}_i,\boldsymbol{\theta}_i)$ is given by 
\begin{equation}
p(\boldsymbol{\Omega}_i,\boldsymbol{\theta}_i|\Phi)=(1-\epsilon)p_S(\boldsymbol{\Omega}_i,\boldsymbol{\theta}_i|\Phi)+\epsilon p_h(\boldsymbol{\Omega}_i,\boldsymbol{\theta}_i|\Phi),
\end{equation}
where $p_S$ is the likelihood given it is a member of the stream as outlined above, and $p_h$ is the likelihood given that it is not a member of the stream (either a member of a smooth background population or another structure). $\epsilon$ is the probability of being an outlier, which is given a logarithmic uniform prior. We choose to specify the outlier model in angle-frequency space as it is simpler to normalize in these coordinates, and we are less sensitive to systematics arising from the St\"ackel approximation. We set $p_h$ to be uniform in both the angle and frequency space, such that
\begin{equation}
p_h(\boldsymbol{\Omega}_i,\boldsymbol{\theta}_i|\Phi) = \frac{\epsilon}{(2\pi)^3 \Omega_{\rm max}^3}
\end{equation}
with $\Omega_{\rm max}=30\Gyr^{-1}$. This model is clearly simplistic, and we are not correctly accounting for selection effects in the angles due to observing stars in some region of the Galaxy.

To our $30$-particle stream dataset we add $20$ outliers from our halo model, such that $\epsilon_{\rm true}=0.4$, and consider the simple case where we have an error-free dataset. The input data-set is shown in Fig.~\ref{galcoords_outlier} and the resulting posterior distributions for the potential parameters are given in Fig.~\ref{errfreeoutlier}. In Fig.~\ref{outlier_frac} we show the posterior distribution for the outlier fraction. It peaks nicely around the input outlier fraction, but has fairly large scatter around this value. This is probably due to the simplistic nature of the background model, but also, as we can see in Fig.~\ref{galcoords_outlier}, there is significant overlap in observable space between the stream and outlier distributions. Further information such as metallicities would potentially provide a simpler way to disentangle the stream from the background.

\section{Discussion}\label{Discussion}
The formalism presented in this paper provides the first steps towards constraining the potential of the Galaxy from a generative stream model expressed in angle-frequency space. Through the course of developing the necessary machinery it has become apparent that many improvements can be made. These are as follows:

(i) The model for a cold stream is very narrow, whilst the error distribution for the observational quantities is expected to be large. The convolution of the error and model distributions only has support over a very narrow range in each observable coordinate. Here we have performed the integration using the sophisticated Monte-Carlo integration method (Vegas) to concentrate rapidly on this small region. However, a more efficient scheme would be full forward modelling. To do this we would perform the Monte Carlo integration by sampling `true' coordinates from the stream distribution and sum the appropriate Gaussian error distributions for each of the observed coordinates. For this we need an efficient scheme for going from a frequency-angle sample to observable space. The torus machine \citep{McMillanBinney2008} was developed to convert from angle-actions to $(\boldsymbol{x},\boldsymbol{v})$ efficiently. Such machinery seems ideal for our purpose. A torus is constructed by finding the coefficients of the generating function from a toy torus. We can construct a series of nearby tori which cover the small range in action space occupied by the stream particles. For a given set of angles we can find a corresponding $(\boldsymbol{x},\boldsymbol{v})$ on each torus. For any action point which lies between tori we can construct an appropriate torus on-the-fly by interpolating the generating function coefficients of the nearby tori. A small complication here is that we wish to work with frequencies instead of actions so a scheme for finding $\boldsymbol{J}(\boldsymbol{\Omega})$ is required. We expect that for the region of interest this function is simple and may be obtained by interpolation between tori. With this function we may also calculate the Hessian as a function of $\boldsymbol{\Omega}$. 

(ii) Our choice of model parametrisation is perhaps more flexible than it should be. All the widths ($w,u,w_0$) are related to $M^{1/3}$ \citep{SandersBinney2013a} so there could be a way to link all three parameters into a single mass parameter. Also, we have allowed the stream to be oriented along some random direction, $\hat{\boldsymbol{n}}$. The direction depends on the exact structure of the action distribution, and the structure of the Hessian matrix. The first of these can be sensitive to internal structure in the progenitor. We gain information about $\hat{\boldsymbol{n}}$ through the use of the angle and frequency structure as discussed in \cite{SandersBinney2013b}. However, we could instead choose to make $\hat{\boldsymbol{n}}$ a function of the progenitor actions in the chosen potential. This would constrain the models in a more physically motivated fashion, and would provide more information when the quality of the data is reduced. 

(iii) We have analysed a simulated stream evolved in a near-spherical logarithmic potential. As discussed in \cite{SandersBinney2013a} this does not exhibit substantial offset between the stream and orbit tracks. Such an offset is more apparent in flattened potentials. Therefore, more tests are required with more realistic potentials with disc components to validate the modelling approach presented here. Additionally, our potential has only two parameters, whilst, in a practical application, a more flexible potential model should be adopted. We have found that the parameters of this two-parameter potential are recovered without significant correlations. From stream data we are measuring the acceleration in some small spatial volume which in the axisymmetric case is given approximately by the gradients $\d\Phi/\d R$ and $\d\Phi/\d z$ at the progenitor. These gradients uniquely specify a combination of $V_c$ and $q$. However, with more parameters in our potential model, many combinations of these parameters would give identical accelerations at the progenitor, so strong correlations are to be expected. More work is required to say exactly what properties of the potential, and hence what potential parameter combinations, a given set of stream data best measures.

(iv) The distributions in frequency and angle, $K_i$, along the principal stream direction, $\hat{\boldsymbol{n}}$, were taken in this paper to be simple Gaussians and a uniform distribution. This procedure is adequate for our purposes but more realistic distributions are required to reproduce the peaky distribution from Fig.~\ref{releasetimes}, and the expected feathering in the stream \citep{Kupper2012,SandersBinney2013b}. For instance, a first step would be to adjust the distribution over stripping times to be more concentrated around pericentric passage of the progenitor by modelling $p(t_i)$ as a comb of Gaussians separated by the radial period of the progenitor model. However, for measuring the Galactic potential the shape of the stream is far more important than the density along the stream and, as we have shown, the assumption of a uniform stripping-time distribution is sufficient for recovery of the potential parameters.

Recently \cite{Bovy2014} presented a machinery very similar to that shown here for constructing models of tidal streams. He exploits the narrow range of frequencies in the stream to construct a simple linear map between $(\boldsymbol{\Omega},\boldsymbol{\theta})$ and $(\boldsymbol{x},\boldsymbol{v})$. The Gaussian structure in $(\boldsymbol{\Omega},\boldsymbol{\theta})$-space can then be simply translated into a Gaussian structure in $(\boldsymbol{x},\boldsymbol{v})$ making marginalization over missing data simpler. \cite{HelmiWhite1999} present a similar approach to analysing a stream model consisting of a Gaussian structure in action space and initial angle space.

\section{Conclusions}
We have presented a probabilistic model for a tidal stream and used this to constrain the potential from a simulation. The stream model builds on the work of \cite{SandersBinney2013b} by constructing a simple model for the stream in frequency and angle space. The presented formalism naturally accounts for the errors in stream data, and can also incorporate the possibility of stream data being contaminated with stars from a smooth halo population or another tidal structure. We have successfully recovered the potential parameters used to run an N-body simulation of a GD-1-like stream from error-free data, data with small errors included, and data with outliers included. 

As currently formulated the computational cost of implementing our approach increases significantly with the magnitude of the observational errors. We have described modifications that promise to mitigate this effect, and thus to make the approach a powerful technique for constraining the Galaxy's gravitational potential.

\section*{Acknowledgements}
JLS thanks James Binney, Hans-Walter Rix, David Hogg, Paul McMillan and the dynamics group in Oxford for useful conversations which shaped this work. JLS acknowledges the support of the Science and Technology Facilities Council, and the comments from the anonymous referee which improved this work.
{\footnotesize{
\bibliographystyle{mn2e-2}
\bibliography{jasonsanders_streams}

\begin{thebibliography}{30}
\expandafter\ifx\csname natexlab\endcsname\relax\def\natexlab#1{#1}\fi

\bibitem[{{Binney}(2008)}]{Binney2008}
{Binney} J., 2008, \mnras, 386, L47

\bibitem[{{Bovy}(2014)}]{Bovy2014}
{Bovy} J., 2014, ArXiv e-prints

\bibitem[{{Bovy} \& {Tremaine}(2012)}]{BovyTremaine2012}
{Bovy} J., {Tremaine} S., 2012, \apj, 756, 89

\bibitem[{{Deg} \& {Widrow}(2014)}]{Deg2014}
{Deg} N., {Widrow} L., 2014, ArXiv e-prints

\bibitem[{{Dehnen}(2000)}]{Dehnen2000}
{Dehnen} W., 2000, \apjl, 536, L39

\bibitem[{{Dehnen}(2002)}]{Dehnen2002}
{Dehnen} W., 2002, Journal of Computational Physics, 179, 27

\bibitem[{{Eyre}(2010)}]{Eyre2010}
{Eyre} A., 2010, DPhil thesis, Univ. Oxford, Oxford

\bibitem[{{Eyre} \& {Binney}(2009{\natexlab{a}})}]{EyreBinney2009B}
{Eyre} A., {Binney} J., 2009{\natexlab{a}}, \mnras, 399, L160

\bibitem[{{Eyre} \& {Binney}(2009{\natexlab{b}})}]{EyreBinney2009A}
{Eyre} A., {Binney} J., 2009{\natexlab{b}}, \mnras, 400, 548

\bibitem[{{Eyre} \& {Binney}(2011)}]{EyreBinney2011}
{Eyre} A., {Binney} J., 2011, \mnras, 413, 1852

\bibitem[{{Foreman-Mackey} {et~al}\mbox{.}(2013){Foreman-Mackey}, {Hogg},
  {Lang}, \& {Goodman}}]{ForemanMackey2013}
{Foreman-Mackey} D., {Hogg} D.~W., {Lang} D., {Goodman} J., 2013, \pasp, 125,
  306

\bibitem[{Galassi {et~al}\mbox{.}(2009)Galassi, Davies, Theiler, Gough, \&
  Jungman}]{GSL}
Galassi M., Davies J., Theiler J., Gough B., Jungman G., 2009, GNU Scientific
  Library - Reference Manual, Third Edition, for GSL Version 1.12 (3. ed.).
  Network Theory Ltd, pp. 1--573

\bibitem[{{Garbari} {et~al}\mbox{.}(2012){Garbari}, {Liu}, {Read}, \&
  {Lake}}]{Garbari2012}
{Garbari} S., {Liu} C., {Read} J.~I., {Lake} G., 2012, \mnras, 425, 1445

\bibitem[{{Helmi} \& {White}(1999)}]{HelmiWhite1999}
{Helmi} A., {White} S.~D.~M., 1999, \mnras, 307, 495

\bibitem[{{Johnston}(1998)}]{Johnston1998}
{Johnston} K.~V., 1998, \apj, 495, 297

\bibitem[{{Johnston} {et~al}\mbox{.}(1999){Johnston}, {Zhao}, {Spergel}, \&
  {Hernquist}}]{Johnston1999}
{Johnston} K.~V., {Zhao} H.~S., {Spergel} D.~N., {Hernquist} L., 1999, in
  Astronomical Society of the Pacific Conference Series, Vol. 194, Working on
  the Fringe: Optical and IR Interferometry from Ground and Space, {Unwin} S.,
  {Stachnik} R., eds., p.~15

\bibitem[{{Koposov} {et~al}\mbox{.}(2010){Koposov}, {Rix}, \&
  {Hogg}}]{Koposov2010}
{Koposov} S.~E., {Rix} H.-W., {Hogg} D.~W., 2010, \apj, 712, 260

\bibitem[{{Kuijken} \& {Gilmore}(1989)}]{KuijkenGilmore1989}
{Kuijken} K., {Gilmore} G., 1989, \mnras, 239, 605

\bibitem[{{K{\"u}pper} {et~al}\mbox{.}(2012){K{\"u}pper}, {Lane}, \&
  {Heggie}}]{Kupper2012}
{K{\"u}pper} A.~H.~W., {Lane} R.~R., {Heggie} D.~C., 2012, \mnras, 420, 2700

\bibitem[{Lepage(1978)}]{Lepage}
Lepage G.~P., 1978, J.Comput.Phys., 27, 192

\bibitem[{{Lux} {et~al}\mbox{.}(2013){Lux}, {Read}, {Lake}, \&
  {Johnston}}]{Lux2013}
{Lux} H., {Read} J.~I., {Lake} G., {Johnston} K.~V., 2013, \mnras, 436, 2386

\bibitem[{{McMillan} \& {Binney}(2008)}]{McMillanBinney2008}
{McMillan} P.~J., {Binney} J.~J., 2008, \mnras, 390, 429

\bibitem[{{Pe{\~n}arrubia} {et~al}\mbox{.}(2012){Pe{\~n}arrubia}, {Koposov}, \&
  {Walker}}]{Penarrubia2012}
{Pe{\~n}arrubia} J., {Koposov} S.~E., {Walker} M.~G., 2012, \apj, 760, 2

\bibitem[{{Price-Whelan} \& {Johnston}(2013)}]{PriceWhelan2013}
{Price-Whelan} A.~M., {Johnston} K.~V., 2013, \apjl, 778, L12

\bibitem[{{Sanders}(2012)}]{Sanders2012}
{Sanders} J., 2012, \mnras, 426, 128

\bibitem[{{Sanders} \& {Binney}(2013{\natexlab{a}})}]{SandersBinney2013a}
{Sanders} J.~L., {Binney} J., 2013{\natexlab{a}}, \mnras, 433, 1813

\bibitem[{{Sanders} \& {Binney}(2013{\natexlab{b}})}]{SandersBinney2013b}
{Sanders} J.~L., {Binney} J., 2013{\natexlab{b}}, \mnras, 433, 1826

\bibitem[{{Tremaine}(1999)}]{Tremaine1999}
{Tremaine} S., 1999, \mnras, 307, 877

\bibitem[{{Varghese} {et~al}\mbox{.}(2011){Varghese}, {Ibata}, \&
  {Lewis}}]{Varghese2011}
{Varghese} A., {Ibata} R., {Lewis} G.~F., 2011, \mnras, 417, 198

\bibitem[{{Zhang} {et~al}\mbox{.}(2013){Zhang}, {Rix}, {van de Ven}, {Bovy},
  {Liu}, \& {Zhao}}]{Zhang2013}
{Zhang} L., {Rix} H.-W., {van de Ven} G., {Bovy} J., {Liu} C., {Zhao} G., 2013,
  \apj, 772, 108

\end{thebibliography}
}}
\appendix
\section{Calculating the Hessian}
As part of our likelihood in Section~\ref{Formalism} we require the Hessian matrix, $\mat{D}=\partial^2H/\partial \boldsymbol{J}^2$. In a St\"ackel potential this matrix may be found numerically by following the procedure presented in the Appendix of \cite{Eyre2010}. This involves finding the second derivatives of the analytic integral expressions for the actions with respect to the integrals of motion: the energy, $E$, the $z$-component of the angular momentum, $L_z$, and the third integral, $I_3$. The resulting integrals are performed analytically using Gaussian quadrature, but care must be taken due to the divergence of the frequency integrand at the end-points. These considerations are taken care of by introduction of a dummy variable as described in \cite{Eyre2010}. 

For our purposes we are using a St\"ackel approximation to the true potential \citep{Sanders2012} so we estimate the true Hessian matrix as that calculated in the approximate St\"ackel potential. In the true potential, the error in each component of the Hessian matrix is less than $10\percent$. However, the error in the determinant is larger ($\sim 30\percent$). As the potentials considered are near-spherical the determinant of the Hessian is small (it is zero for the spherical case), which arises due to cancelling terms in the calculation. Therefore, small errors in each component can give rise to larger errors in the determinant. We recover the appropriate trends in the determinant of the Hessian matrix with the potential parameters. There is a slight bias in Fig.~\ref{errfree} which is probably due to the errors in the Hessian matrix. However, it is not significant. The results shown in this paper demonstrate that the magnitude of the observational errors in the data dominates any systematic errors in estimating the angles, frequencies, or the determinant of the Hessian matrix.

A better, but more time-consuming, estimate of the Hessian matrix may be found using the torus machine \citep{McMillanBinney2008} as described in \cite{SandersBinney2013a}.

\label{lastpage}

\bsp
\end{document}